\documentstyle[aps,prl,multicol,floats,epsf]{revtex}

\begin{document}
\draft
\wideabs{

\title{Mobility of the Doped Holes and the Antiferromagnetic 
Correlations in \\
Underdoped High-$T_c$ Cuprates}

\author{Yoichi Ando, A. N. Lavrov, Seiki Komiya, Kouji Segawa, 
and X. F. Sun}
\address{Central Research Institute of Electric Power 
Industry, Komae, Tokyo 201-8511, Japan}

\date{\today}
\maketitle

\begin{abstract}
The emergence and the evolution of the metallic charge transport 
in the La$_{2-x}$Sr$_x$CuO$_4$ system from lightly- to 
optimally-doped samples ($x$ = 0.01 -- 0.17) are studied.
We demonstrate that in high-quality single crystals the in-plane 
resistivity shows a metallic behavior for {\it all} values of $x$ 
at moderate temperatures and that the hole mobility at 300 K 
changes only by a factor of 3 from $x$=0.01 to 0.17, where its 
$x$-dependence is found to be intriguingly similar to that of the 
inverse antiferromagnetic correlation length.
We discuss an incoherent-metal picture and a charged-stripe scenario 
as candidates to account for these peculiar features.
\end{abstract}

\pacs{PACS numbers: 74.25.Fy, 74.25.Dw, 74.20.Mn, 74.72.Dn, 74.72.Bk}
}
\narrowtext

The underdoped region of the high-$T_c$ cuprates has 
attracted significant attention in recent years. 
Quite peculiar features such as pseudogap \cite{Timusk}, 
$\log (1/T)$ insulating behavior \cite{logT}, 
and charged-stripe instability \cite{Tranquada,Yamada}
have been discovered in the underdoped region, 
and those features are expected to bear essential clues to 
elucidating the origin of the high-$T_c$ superconductivity.
The La$_{2-x}$Sr$_x$CuO$_4$ (LSCO) system has been most 
frequently used for systematic studies of the moderately- to 
heavily-underdoped region of the cuprates, because of the 
accessibility to the low-doped region.
Neutron scattering and photoemission 
experiments on the LSCO system 
have provided detailed knowledge about the evolution of the 
spin \cite{Yamada,Kastner} and electronic \cite{Ino1,Ino2} 
structures as holes are 
doped to the parent Mott insulator; however, how the 
metallic charge conduction emerges in this doped Mott insulator 
is still poorly understood. 
It has been proposed theoretically that, because of the 
strong magnetic correlations in cuprates, 
the metallic conduction may be realized through 
spin-charge separated quasiparticles \cite{Anderson,Lee}, 
charged stripes \cite{Emery,Zaanen}, 
incoherent quasiparticles \cite{Georges}, {\it etc.}
To fully understand the transformation from the Mott 
insulator to a metal in cuprates, systematic studies of 
the transport properties in the heavily-underdoped region are 
highly desirable.

In the early days of high-$T_c$ research, the transport 
properties of the heavily-underdoped LSCO were studied 
using polycrystals, thin films, 
and flux-grown crystals \cite{Kastner}.  
These works showed that the charge transport became 
insulating in the non-superconducting samples with the 
variable-range-hopping (VRH) conductivity at low temperatures, 
while at high temperatures the transport could 
be metallic already in the spin-glass regime \cite{Kastner}.  
However, there has been no systematic 
measurement of the transport properties of the high-quality 
single crystals in the heavily-underdoped region, which have 
become available with the development of the traveling-solvent 
floating-zone (TSFZ) technique \cite{FZ}.  
Since one can achieve high uniformity of the Sr concentration 
and high purity in the TSFZ-grown crystals, it has become 
possible to study, for example, the doping dependence of the 
mobility of the holes in clean and well-controlled crystals 
down to the heavily-underdoped antiferromagnetic region 
without being bothered by extraneous disorder.

In this Letter, we report the in-plane resistivity $\rho_{ab}$ 
and the Hall coefficient $R_H$ measurements of a series of 
TSFZ-grown high-quality LSCO crystals that cover the whole 
underdoped region, from $x$=0.01 to 0.17. 
It is found that in high-quality crystals the behavior of 
$\rho_{ab}(T)$ is metallic ($d\rho_{ab}/dT > 0$) at  
moderate temperatures for all values of $x$, 
even in the N\'{e}el state at $x$=0.01; since the magnitude 
of $\rho_{ab}(T)$ in the lightly-doped region are far above 
the Mott limit for metallic conduction in two-dimensional (2D)
metal, this is a strong manifestation of the ``bad metal" 
\cite{Bad_Metal} behavior. 
Moreover, we observe that the mobility of doped holes in this 
``metallic" regime shows a doping dependence similar 
to that of the inverse antiferromagnetic (AF) correlation length, 
$\xi_{\rm AF}^{-1}$.  Notably, the absolute 
value of the mobility changes only by a factor of 3 from 
$x$=0.01 to 0.17 at 300 K. 
These results demonstrate the peculiar nature of the 
charge transport in the cuprate that is a doped Mott insulator 
with strong magnetic correlations, and give essential clues to 
elucidate the origin of the bad metal behavior in cuprates.
To corroborate the result on LSCO, we also show the $\rho_{a}(T)$
data of YBa$_2$Cu$_3$O$_y$ (YBCO) for similarly wide range 
of doping.

The clean single crystals of LSCO are grown by the TSFZ technique 
and are carefully annealed to remove excess oxygen, 
which ensures that the hole doping is exactly equal to $x$.  
The crystallographic axes are determined by the Laue 
analysis and then the samples are shaped into thin 
platelets with the $ab$ planes parallel to the wide face, 
where an error in the axes directions is 
less than 1$^{\circ}$.
The YBCO crystals are grown in Y$_2$O$_3$ crucibles by a 
conventional flux method and are detwinned at temperatures 
below 220$^{\circ}$C with an uniaxial pressure of $\sim$0.1 GPa 
after the oxygen content is tuned to the desired doping with 
careful annealing and quenching \cite{Segawa_60K}.
The resistivity in YBCO is measured along the $a$-axis 
to exclude the conductivity contribution from CuO chains, 
which run along the $b$-axis.
Note that the detwinning is not necessary for YBCO crystals 
with $y < 6.40$, which have tetragonal symmetry.

The in-plane resistivity $\rho_{ab}$ and the Hall coefficient 
$R_H$ are measured using a standard ac six-probe method. 
The Hall effect measurements are done by sweeping the 
magnetic field to $\pm$14 T at fixed temperatures stabilized 
within $\sim$1 mK accuracy \cite{Ando}. 
The uncertainty in the absolute magnitude of $\rho_{ab}$ and 
$R_H$ is minimized \cite{Segawa_60K} by using relatively 
long samples (voltage 
contact separation is typically $\sim$1 mm for YBCO and 
$\sim$1.5 mm for LSCO), painting narrow contact pads with the 
width of 50--80 $\mu$m, and accurately determining the crystal 
thickness by measuring the weight with 0.1-$\mu$g resolution; 
total errors in absolute values of $\rho_{ab}$ and $R_H$ are 
less than 10\% and 5\%, respectively. 
In our crystals, $\rho_{ab}$ is very reproducible 
(we always measure several crystals for each composition) 
and its absolute value is among the smallest ever reported 
for each composition.

\begin{figure}[t!]
\epsfxsize=0.85\columnwidth
\centerline{\epsffile{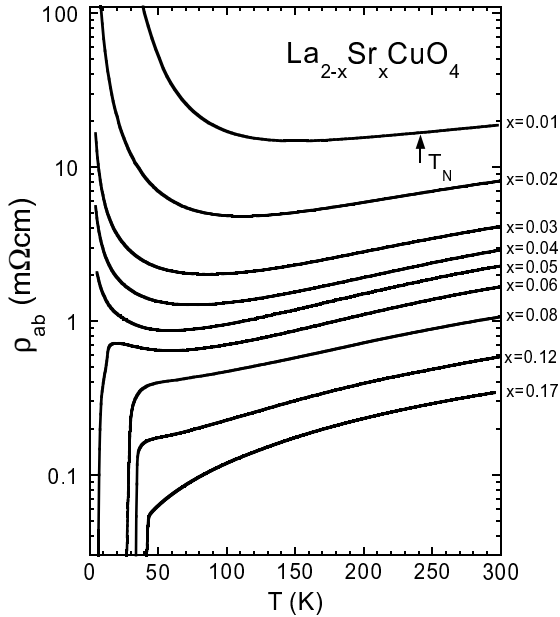}}
\vspace{0.2cm}
\caption{Temperature dependences of 
$\rho_{ab}$ in TSFZ-grown La$_{2-x}$Sr$_x$CuO$_4$ single crystals.}
\label{fig1}
\end{figure}

Figure 1 shows the temperature dependences of $\rho_{ab}$ for 
the LSCO crystals, with the vertical axis in the logarithmic scale. 
One may immediately notice that $\rho_{ab}$ in the 
moderate-temperature range show metallic behavior 
($d\rho_{ab}/dT > 0$) for {\it all} values of $x$.
It is particularly intriguing to note that in the $x$=0.01 sample 
$\rho_{ab}(T)$ keeps its ``metallic" behavior 
well below the N\'{e}el temperature $T_N$ 
(which is 240 K, see Fig. 2 inset).  This clearly demonstrates 
that the in-plane charge transport is insensitive 
to the long-range magnetic order, which may not be surprising 
because the large $J$ ($\sim$0.1 eV) causes the 
antiferromagnetic correlations to be well established in the 
CuO$_2$ planes far above $T_N$ \cite{Kastner}.

\begin{figure}[t!]
\epsfxsize=0.8\columnwidth
\centerline{\epsffile{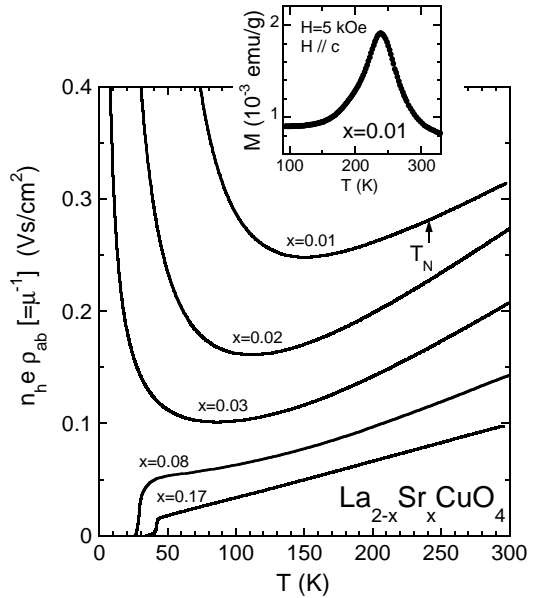}}
\vspace{0.2cm}
\caption{Temperature dependences of the inverse mobility, 
$n_{h}e\rho_{ab}$, of the LSCO crystals.  
Inset: Magnetization of a large La$_{1.99}$Sr$_{0.01}$CuO$_4$ 
single crystal (93 mg) from which the samples for $\rho_{ab}$ 
measurements were cut; the peak in $M(T)$ corresponds to the 
N\'{e}el transition.}
\label{fig2}
\end{figure}

To examine the detailed doping dependence of the charge transport,
it is useful to look at the conductivity per charge, namely 
the mobility of the doped holes.
In Fig. 2, we plot the temperature dependences of 
$n_{h}e\rho_{ab}$ for various $x$,
where $e$ is the electronic charge and $n_h$ is the nominal hole 
concentration given by $2x/V$ 
[unit cell $V$ ($\simeq 3.8\times 3.8\times 13.2$ \AA$^3$) 
contains two CuO$_2$ planes in LSCO].  
This product, $n_{h}e\rho_{ab}$, corresponds to the inverse mobility 
$\mu^{-1}$ of the doped holes.  
Note that it is probably better to use $n_{h}e$ (= $2ex/V$) 
than to use $R_H^{-1}$ for the calculation of the conductivity 
per hole, because $R_H$ in cuprates shows a strong temperature 
dependence that is not caused by a change in the density of 
mobile holes.
One can see in Fig. 2 that the slope of 
$n_{h}e\rho_{ab}(T)$ at 300 K depends only weakly on $x$ 
and the absolute value of $\mu^{-1}$ 
changes only by a factor of 3 at 300 K; the similarity of the 
moderate-temperature $\mu^{-1}(T)$ curves and the rather small 
change in their magnitude hint at the 
possibility that the ``metallic" charge transport is governed by 
essentially the same mechanism from $x$=0.01 to 0.17.

\begin{figure}[t!]
\epsfxsize=0.85\columnwidth
\centerline{\epsffile{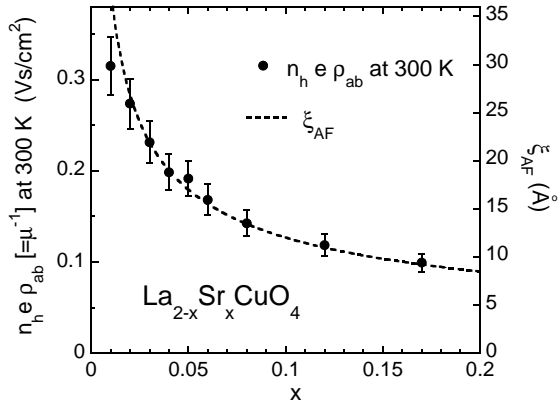}}
\vspace{0.2cm}
\caption{$x$ dependence of the inverse mobility (left axis) 
of the LSCO crystals.  The dashed line shows the $x$ dependence 
of $\xi_{\rm AF}$ (right axis), which is reported to be 
$3.8/\sqrt{x}$ \AA \ \ by neutron experiments 
\protect\cite{Kastner}.}
\label{fig3}
\end{figure}

Figure 3 shows the $x$ dependence of $\mu^{-1}$ at 300 K.
The change in $\mu^{-1}$ is smooth and is relatively small, 
which is rather surprising in view of the 
superconductor-insulator transition occurring at 
$x\approx$ 0.05.  In Fig. 3, we also plot the 
$x$ dependence of the AF correlation length $\xi_{\rm AF}$, 
which is known to show similarly smooth change with $x$; 
$\xi_{\rm AF}(x)$ is reported to be described by 
$3.8/\sqrt{x}$ \AA \cite{Kastner}, 
the average separation between the doped holes. 
(Note that $\xi_{\rm AF}$ is almost temperature independent 
below $\sim$300 K except for $x$=0.01 \cite{Kastner}.) 
It is striking that both $\mu^{-1}$ and $\xi_{\rm AF}$ 
show very similar $x$ dependence; in particular, both show 
only a factor of 2 or 3 change from $x$=0.02 to 0.17 despite 
the significant change in the ground state.
The similarity of the $x$ dependences of $\mu^{-1}$ and 
$\xi_{\rm AF}$ suggests that the charge transport 
in the underdoped cuprates is inherently related 
to the magnetic correlations in the background spins. 
It is worthwhile to mention that the magnitude of $\mu$ 
of LSCO is 3 -- 10 cm$^2$/Vs at 300 K in the whole doping 
range studied, and these numbers are close to those of 
typical metals (e.g. $(ne\rho)^{-1}$ of iron is 
4 cm$^2$/Vs at 273 K \cite{Ashcroft}). 

We note that the ``metallic" behavior of $\rho_{ab}(T)$ at 
moderate temperatures observed in the heavily-underdoped 
samples ($x \le 0.05$) indicates that the system is not 
an insulator (or a semiconductor) 
with a well-defined gap.  The low temperature insulating 
behavior in these samples is consistent with the 
VRH behavior as reported before \cite{Kastner}, 
which also suggests that there is a developing band of 
electronic states at the Fermi energy $E_{\rm F}$.
Therefore, given that the chemical potential is pinned in the 
middle of the Mott-Hubbard gap \cite{Ino2}, 
the resistivity data tell us that with only 1\% of 
doping a band is created near $E_{\rm F}$ within the 
gap of the parent cuprate, and 
the system starts to show band-like transport (with disorder). 
The Hall coefficient data also support this picture; namely, 
like in ordinary metals, the apparent hole density 
$n \equiv (eR_{H})^{-1}$ of the heavily-underdoped samples is 
essentially temperature independent in the temperature range 
where the metal-like behavior of $\rho_{ab}(T)$ is observed, 
as shown in Fig. 4 for $x$=0.01 -- 0.03.  
Note that, at low doping, $n$ agrees well with the nominal 
hole density $n_h$ at moderate temperatures, 
which means that all the doped holes are moving and 
contributing to the Hall effect.  

The above results on LSCO are essentially reproduced in YBCO.
Figure 5 shows the temperature dependences of 
$n_{h}e\rho_{a}$ for clean untwinned YBCO crystals with 
various oxygen contents.  
Although the determination of hole concentrations in YBCO is 
not as straightforward as in LSCO, one may estimate \cite{Hanaki} 
the hole number per Cu to be around 0.02, 0.05, and 0.17 
for samples with $y = 6.30$, 6.45, and 6.95, respectively.  
These doping levels cover the range from the AF region 
($T_N = 230$ K for $y = 6.30$) to optimally-doped superconductor 
with $T_c = 93$ K.  
As is the case with LSCO, the metallic behavior is insensitive to 
the establishment of the AF state at $T_N$, 
and $\mu^{-1}$ at 300 K changes only by a factor of 2 
from the AF ``insulator" to the 
optimally-doped superconductor.  
Note that both LSCO and YBCO show similar $\mu$ of around 
10 cm$^2$/Vs at moderate temperatures, which suggests that 
the hole mobility is almost universal among the cuprates. 
Also, though not shown here, the $R_H(T)$ data of the 
lightly-doped YBCO are essentially similar to those 
of LSCO and are consistent with the data published recently 
\cite{Semba}. 

\begin{figure}[t!]
\epsfxsize=0.8\columnwidth
\centerline{\epsffile{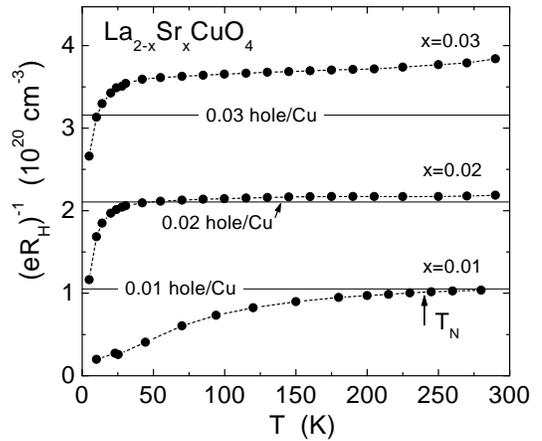}}
\vspace{0.2cm}
\caption{The apparent density of carriers $n = (eR_H)^{-1}$ 
for the three lightly-doped LSCO crystals; 
solid lines indicate the nominal hole density $n_h$ for 
each doping.}
\label{fig4}
\end{figure}

In the above results, one of the notable findings is that 
a ``metallic" behavior 
of $\rho_{ab}(T)$ at moderate temperatures is already 
established even for 1\% of hole doping.  
Although such a metallic behavior 
suggests a Boltzmann band transport for the doped holes,
the {\it magnitude} of $\rho_{ab}$ is so large that 
$k_{\rm F}l$ would be only 0.1, which strongly violates 
the Mott limit for the metallic transport 
($k_{\rm F}$ is the Fermi wave number and $l$ is the 
mean free path) \cite{Mott_Limit}, and therefore the charge 
transport cannot be caused by the motion of ordinary 
quasiparticles in a 2D electronic system.  
In this sense, this is a strong manifestation of the 
``bad metal" \cite{Bad_Metal} behavior at low doping, 
although the bad metals have often been discussed 
in conjunction with the absence of resistivity saturation 
at high temperature \cite{Bad_Metal}. 
Our results on the behavior of $\mu^{-1}$ (Figs. 2 and 3) 
suggests that the mechanism that causes the bad metal behavior 
in cuprates is likely to be fundamentally related to the 
background AF correlations and that the same mechanism may 
well govern the charge transport up to $x$=0.17.
Below we discuss two possibilities for the charge transport 
mechanism that may be consistent with the peculiarities 
observed here.

\begin{figure}[t!]
\epsfxsize=0.8\columnwidth
\centerline{\epsffile{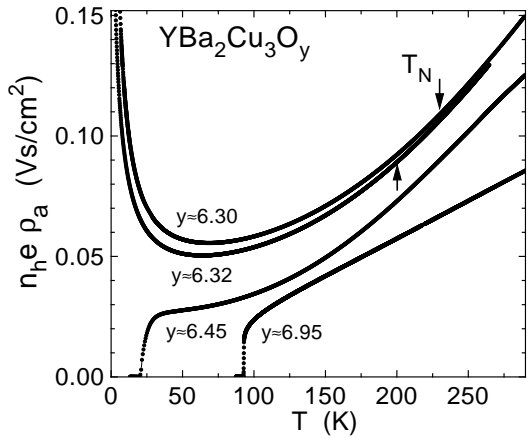}}
\vspace{0.2cm}
\caption{Temperature dependences of the inverse mobility, 
$n_{h}e\rho_{a}$, of untwinned YBa$_2$Cu$_3$O$_y$ crystals.  
The N\'{e}el temperatures for lightly-doped 
($y = 6.30$ and 6.32) crystals are indicated by arrows.}
\label{fig5}
\end{figure}

One possible picture is the ``incoherent metal" 
\cite{Ino1,Georges}, 
where $E_{\rm F}$ is smaller than $k_{\rm B}T$ and thus 
the charge transport is essentially a diffusion process 
of classical particles; since the charge dynamics is 
diffusive and thus is not strictly a band transport, 
the Mott limit is completely irrelevant in this case.  
Still, the peculiar $T$-linear resistivity may be 
expected in an incoherent metal \cite{Georges}, 
because the large magnetic energy scale causes the system 
to be quantum critical.
Recent photoemission results on LSCO 
also suggest that the incoherent metallic state 
may emerge in the lightly-doped region from the parent 
Mott insulator \cite{Ino1}.
However, it may be difficult to understand the apparent 
connection between $\mu^{-1}$ and $\xi_{\rm AF}$ 
in the incoherent-metal picture.

Another possibility is that the phase segregation on the 
mesoscopic scale changes the effective volume relevant to 
the charge transport. 
In this case, both the apparent violation of the Mott 
limit and the insensitivity of the mobility to the change in 
$x$ can be naturally understood.
For example, when the charged stripes are formed, 
metallic transport through the quasi-one-dimensional 
charge system may be possible \cite{Emery,Zaanen}. 
Since the charged stripes are rather decoupled from the 
magnetically-ordered regions that separate them, 
the metallic behavior may only weakly depend on $x$; 
however, the apparent correlation 
between $\mu^{-1}$ and $\xi_{\rm AF}$ indicates 
that the charge transport is influenced by the rigidity 
of the magnetic correlation in the magnetic domains,
which probably means that the transverse fluctuation of the 
stripes in the AF environment must be significant within 
this picture. 
In fact, a recent theory predicts \cite{Kivelson} that the 
transverse fluctuation of the charged stripes gives rise to 
an ``electronic liquid crystals", which can behave like a 
2D metal.

In summary, 
we show that the inverse mobility $\mu^{-1}$ at moderate 
temperatures shows a metallic temperature dependence in the 
whole underdoped region down to $x$=0.01 and that the magnitude 
of $\mu^{-1}$ at 300 K changes only by a factor of 3 from $x$=0.01 
to 0.17 in high-quality single crystals.
Moreover, it is found that the $x$-dependence of $\mu^{-1}$ at 
300 K appears to be intriguingly similar to that of $\xi_{\rm AF}$ 
in the whole underdoped region.
These features suggest that the {\it same} mechanism govern 
the charge transport from $x$=0.01 to 0.17 and that such mechanism 
is fundamentally related to the background AF correlations. 
It is discussed that both an incoherent metal or charged stripes 
can basically explain the unusual ``metallic" charge transport, 
though the apparent correlation between $\mu^{-1}$ 
and $\xi_{\rm AF}$ can be more easily understood within the 
stripe scenario.

We would like to thank A. Fujimori for helpful discussions, 
and T. Sasagawa and K. Kishio for valuable suggestions for LSCO 
crystal growth.

%
\medskip
\vfil


\begin{references}
\bibitem{Mott_Limit}
In 2D systems $k_{\rm F}l$ can be calculated by 
$hc_0/\rho_{ab}e^2$, where $c_0$ is the interlayer distance. 
When $k_{\rm F}l < 1$, $l$ is shorter than the 
de Broglie wavelength of the electrons at $E_{\rm F}$, 
and thus {\bf k} is not well-defined; 
this means that the band picture breaks down \cite{Bad_Metal}.
\end{references}
\end{document}